\documentclass[runningheads]{svmult}

\usepackage{makeidx}   
\usepackage{graphicx}  
\usepackage{subeqnar}  
\usepackage{multicol}  
\makeindex             

\def\lesssim{\mathrel{\hbox{\rlap{\hbox{\lower4pt\hbox{$\sim$}}}\hbox{$
<$}}}}
\def\gtrsim{\mathrel{\hbox{\rlap{\hbox{\lower4pt\hbox{$\sim$}}}\hbox{$>
$}}}}
\begin{document}
\title*{Jet Formation and Collimation in AGN and $\micro$-Quasars}
\toctitle{Jet Formation and Collimation
\protect\newline }
%
%
\titlerunning{Jet formation and collimation}
%
\author{Christophe Sauty\inst{1}
\and Kanaris Tsinganos\inst{2}
\and Edoardo Trusssoni\inst{3}}
\authorrunning{C. Sauty et al.}
\institute{
Universit\'{e} Paris 7 - Observatoire de Paris, D.A.E.C.,
F-92190 Meudon, France
\and 
Department of Physics, University of Crete, 
GR-710 03 Heraklion, Crete, Greece.
\and
Osservatorio Astron. di Torino, Strada
Osservatorio 20, I-10025 Pino Torinese, Italy
}

\maketitle              

\begin{abstract}
We briefly review our current understanding for the formation, 
acceleration and  collimation of winds to jets associated with compact 
astrophysical objects such as AGN and $\micro$Quasars. 

All such outflows may be considered to a first approximation as ideal 
MHD plasmas escaping from a rotating and magnetized accretion disk with a   
magnetosphere around a central black hole. A crucial ingredient for a correct 
modelling of the steady state problem is to place the appropriate boundary 
conditions, by taking into account how information can propagate through 
the outflow and ensuring, e.g., that shocks produced via the interaction 
of the  flow with the external medium do not affect the overall structure. 
As an example underlining the role of setting the correct boundary 
conditions, we make the analogy of the critical surfaces in the steady and 
axisymmetric MHD problem with the event horizon and ergosphere of a 
rotating black hole in relativity.

We discuss the acceleration of the outflow, by gas, radiation, or wave  
pressure gradients and also by magnetic mechanisms, showing the important 
role played by the disk corona in the vicinity of the black hole.
Pressure and magnetic confinement both may also play a role in 
confining the outflow, although magnetic hoop stress confinement is likely to 
be a rather dominant process in tightly collimated outflows. 
The possible asymptotical morphology that jets achieve and the instabilities which 
are likely to explain the observed structures but do not prevent jets to 
possess toroidal magnetic fields are also reviewed.

Finally, it is proposed that in a space where the two main variables are the 
energy of the magnetic rotator and the angle between the line of sight and 
the ejection axis,  some observed characteristics of AGN jets can be 
understood. A criterion for the transition of the morphologies of the outflows 
from highly collimated jets to uncollimated winds is given. It is based on the 
analysis of a particular class of exact solutions and may somehow generalize 
other earlier suggestions, such as the spinning of the black hole, the 
fueling of  the central object, or the effects of the environment. 

Thus, while  the horizontal AGN 
classification from Type 0 to Types 1 and 2 may well be an orientation effect --
i.e. a dependence on the viewing angle between the source axis and the observer
as in the standard model -- the vertical AGN classification with 
uncollimated outflows (radio-quiet sources) and collimated outflows 
(radio-loud sources) depends both, on the efficiency 
of the magnetic rotator and the environment in which the outflows propagate. 

\end{abstract}


\section{Introduction} 

\subsection{Schematic picture of AGN}

\begin{figure}[b]
\begin{center}
\includegraphics[width=.5\textwidth]{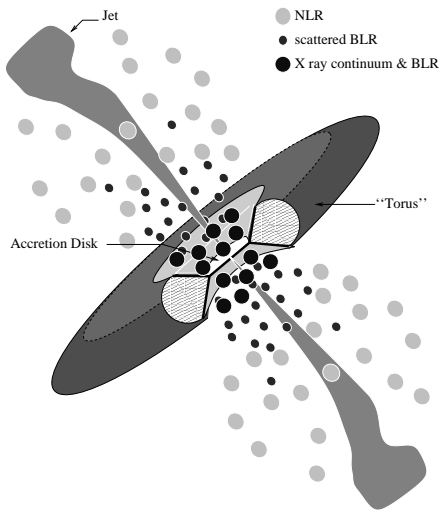}
\end{center}
\caption[]{General sketch, not to scale, of an AGN following Urry and Padovani 
(\cite{UP95}, see the text for details).}
\label{eps1}
\end{figure}

Some galaxies are known to emit radiation with extremely high luminosities 
from rather small volume in  the $\gamma-$ray, X-ray and UV continuum.   
Such active cores are the so-called Active Galactic Nuclei (AGN) and the radiation 
 is commonly believed to be gravitational energy released by matter spiraling 
around a supermassive central black hole of about 10$^9M_{\odot}$  (see Fig. 
\ref{eps1}). 

Though the central engine which produces the enormous observed 
activity cannot be resolved observationally, a standard picture of an AGN
has gradually emerged to explain the richness of the radiation spectra:
\begin{itemize}
\item{} an accretion disk from about 2 to 100 gravitational radii, $R_g$, 
feeding 
the central black hole and emitting mainly in the UV and soft X-rays;
\item{} the broad line optically emitting clouds (BLR), which seem to be absent 
in some sources (e.g. FRI, see 
hereafter) and extend up to a few  $10^3R_g$ from the center. The BLR emission 
can be radiation scattered by hot electrons further away  while the word ``cloud'' should be 
taken in the broad sense meaning  
dense gas with a filling factor less than unity \cite{PietriniTorricelli}; 
\item{} a dusty torus (or wrapped disk or dusty bipolar flow)
with an inner radius of a few $10^3 R_g$, which obscures the central parts of the AGN
from transverse lines of sight; 
\item{} the narrow line regions (NLR) 
which extend from about $10^4$ to  $10^6 R_g$; 
\item{}  powerful jets of plasma detected from the sub-parsec to the 
Mpc scales, mainly in the radio but also in the optical, UV and X-rays. 
\end{itemize}
Note also, that ultra high energy $\gamma$-rays have also been observed from the central 
regions of several Blazars. Jets, together with the emission of radiation from the
immediate neighborhood of an AGN provide a crucial link between the 
easier observed large Mpc scale 
and the sub-parsec scales where presumably the plasma of the jets is  
accelerated in a few gravitational radii from the center of the AGN.  

\subsection{Unified Schemes for AGN}
\begin{figure}[b]
\begin{center}
\includegraphics[width=.7\textwidth]{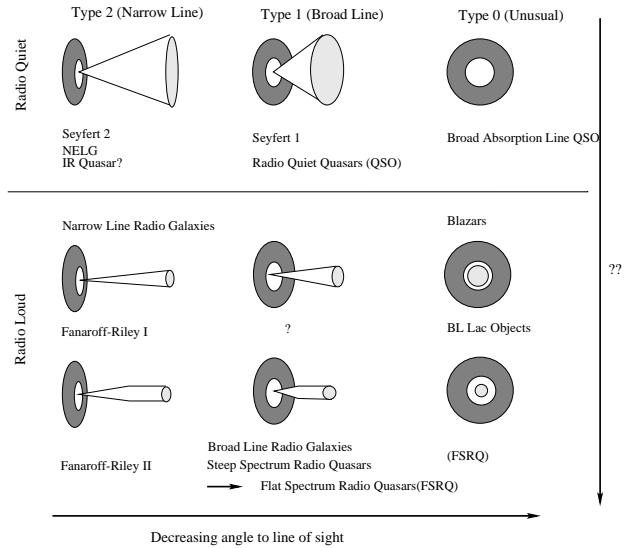}
\end{center}
\caption[]{Unified scheme presented by Urry and Padovani 1995  (\cite{UP95}). Properties
of AGN depend on at least two parameters: the viewing angle and some other 
parameter yet to be defined. }
\label{eps2}
\end{figure}

Based on the phenomenology of their emission in the radio and optical/UV 
parts of the spectrum, AGN are commonly divided into three broad classes, 
as it may be found in excellent reviews in \cite{Antonucci93,UP95}. 
Although some details have been already modified since then, the overall 
classification still holds nowadays, at least as far as the properties of the
associated winds and jets are concerned:
\begin{itemize}
\item \underbar{Type 2} AGN 
have weak continua with narrow emission lines (NLR). They include, 
in the radio quiet group, the low luminosity Seyfert 2 galaxies and narrow 
emission line galaxies (NELG) while the radio-loud counterpart regroups the 
narrow-line radio galaxies with the two distinct morphologies of Fanaroff-Riley I 
(low-luminosities, FR I) and Fanaroff-Riley II (higher luminosities,
 FR II).
\item\underbar{Type 1} AGN 
have bright continua with broad emission lines (BLR) in addition 
to the NLR. The type 1 radio-quiet group is composed of the low luminosity 
Seyfert 1 galaxies (Sey 1) and the higher luminosity radio-quiet quasars 
(QSO), while the radio-loud group includes the broad line radio galaxies 
(BLRG) at low luminosities and the flat or steep spectrum radio-loud  quasars
 (FSRQ and SSRQ) at higher luminosities. 
\item\underbar{Type 0} AGN 
correspond to the remaining AGN with weak or unusual emission 
lines, i.e., in the radio-quiet end the broad absorption lines QSO (BALs) and 
in the radio-loud end the Blazars (BL Lacs and flat spectrum radio quasars, 
FSRQ).
\end{itemize}
Thus, the various AGN can be classified according to orientation, 
beaming and obscuration effects \cite{UP95}.
In this classification scheme, the transition from Type 0 to Type 1 and then 
to Type 2 of the class of radio-quiet AGN is based on orientation effects 
alone (Fig. \ref{eps2}). Namely, in Type 0 AGN the line of sight is almost 
coincident with the axis of the 
system, while the disk is seen face-on. In Type 2 the viewing angle is close 
to 90$^o$ (disk edge-on) and obtains intermediate values for Types 1. The 
broad emission lines arise from ``clouds'' (i.e. dense gas with a small filling 
factor)
orbiting above but nearby the disk (Fig. \ref{eps1}). Thus when the line of 
sight makes a small angle with the system axis, they are not obscured by the
dusty torus, as in  Seyfert~1, while wherein these broad emission lines are
obscured by the torus, only the narrow emission lines are visible because
they are produced further away, as it is the case with Seyfert~2. 

Similarly among radio-loud galaxies, the transition from Type 0 (Blazars) to 
Type 2 (FR I/II radio galaxies) is based on a combination of orientation with 
relativistic beaming, i.e., whether a radio-loud AGN is a radio galaxy or a 
Blazar, and also depends on the angle between its relativistic jet 
and the line of sight. In this sense there seems to 
be a transition from FSRQ (Type 0) to SSRQ (Type 1) and then FR II (Type 2). 
For low luminosity radio loud galaxies there seems to be a gap as BL Lac objects (Type 0) 
are associated with FR I (Type 2) with no Type 1 counterpart. Although this 
association is still controversial,
it may be explained by an intrinsic absence of  broad emission line clouds 
\cite{Chiab99} which would prevent to find any corresponding 
Type 1 objects with a BLR. This argument is supported by the fact that with increasing  
resolution BLR are also sometimes detected in FR II.
At the same time however, recent data at optical and X-ray wavelengths  
have shown that, for the FR I/BL Lac case, the standard unification 
model does not seem to be in full agreement with observations.
A possible way out to  reconcile observations with the standard 
unification scheme  is to assume a structure of the
velocity across the jet  (\cite{Capettietal00,Chiabergeetal00}).   
  
It is now clear that orientation effects are not sufficient to explain the
 difference between radio-quiet, low luminosity and radio loud and high 
luminosity galaxies and quasars. It seems that in radio-loud  
AGN the outflow is relativistic at least in parsec scales, very well 
collimated in the form of a jet and quite powerful on large scales 
where it feeds the terminal radio lobes. 
Conversely, in radio-quiet AGN the outflow is either stopped or 
loosely collimated in the form of a wind or a bipolar flow. Parallely
FR II jets are much more powerful than in FR I with a higher degree of
collimation and terminal hot spots. Simultaneously the environment of the 
jets in FR I sources seems richer than in FR II ones.
Various possibilities have been suggested. The radio-loudness could be 
related to: 
(i) the host galaxy type \cite{Smithetal86},
(ii) the black hole's spin  separating the  lower spin radio-quiet galaxies from the higher black 
hole spin in radio-loud galaxies \cite{Blandford90,WilsonColbert95}, 
(iii) the differences in the rate of nuclear feeding \cite{Reesetal82,Baumetal95}, 
(iv) the  different composition of the plasma \cite{Celotti98}, or
 (v)  the different interaction with the ambient medium \cite{GopalWiita00}. 
Nevertheless,  none of these scenarios seem to be completely satisfactory
because for all of them counter-examples may be found. We suggest at the end of this 
review a quantitative physical criterion for such a transition from 
radio-quiet to radio-loud galaxies which may, in fact, reconcile those various 
points of view by taking a different approach.

\begin{figure}[b]
\begin{center}
\includegraphics[width=.6\textwidth]{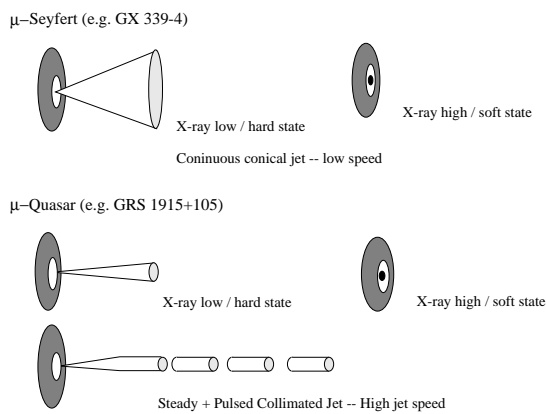}
\end{center}
\caption[]{Summary of the outflow properties of the galactic counterparts of 
AGN  \cite{Fender00}. See text for details.}
\label{eps3}
\end{figure}
\subsection{Towards a similar unification scheme for $\micro$Quasars ?}

The galactic counterparts of the extra-galactic AGN were 
discovered recently by Mirabel and collaborators 
(\cite{MirabelRodriguez99} and references therein). 
Although the central black hole is not supermassive but just of the order of 
one solar mass, $M_\odot$, they also have relativistic ejecta with similar beaming 
effects. It is of course too early to draw a precise classification of such 
objects, since their number is rather small in comparison to AGN. Nevertheless,
there seem to exist $\micro$-Seyferts and $\micro$-quasars (\cite{Corbeletal00,Fender00} and 
Fender's review in this volume for details), with  
prototypes GX 339-4 and GRS 1915+105, respectively (Fig. \ref{eps3}). 
The winds of $\micro$-Seyferts seem to be more continuous and conical while
$\micro$-quasars seem to have steady outflows in addition to pulsed 
collimated jets with higher speed (\cite{Dhawanetal00} and references therein). 
However, these objects show also in X-rays low/hard states where the 
ejection is present and a high/soft state where no outflow is produced, 
probably because of the disruption of the disk in the immediate vicinity of 
the central black hole. 
It is also interesting to note that especially in GX 339-4, the presence of 
the wind is associated with an extended X-ray corona at its base.

Note that we do not include in the present discussion all galactic 
relativistic jets from other binary systems but only those which have similar 
properties with AGN.  
Nevertheless, most of the mechanisms reviewed here apply 
also to such jets as they also apply, incidentally, to jets from young stars,
stellar winds, etc. This may explain why the application of the theory of 
MHD winds, in jets from Young Stars and AGN has evolved parallely. 

\begin{figure}[b]
\begin{center}
\includegraphics[width=.85\textwidth]{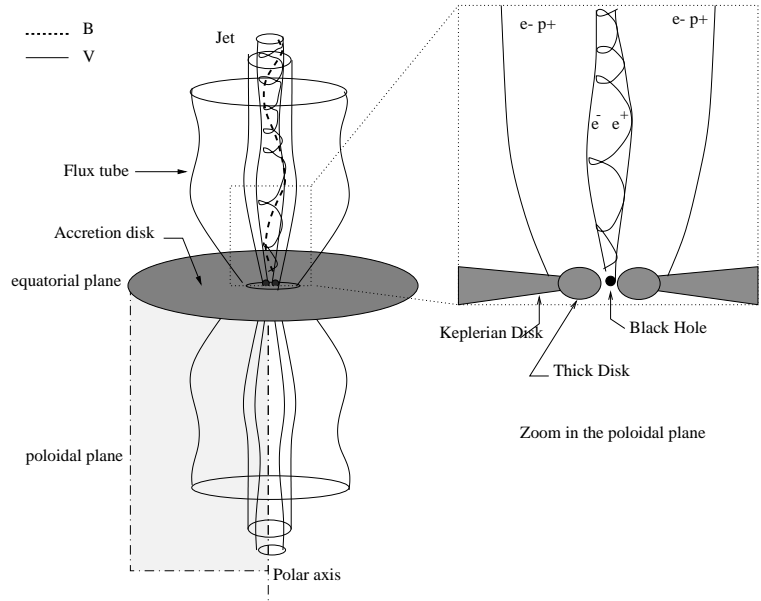}
\end{center}
\caption[]{View of the jet formation region. 
}
\label{eps4}
\end{figure}
\subsection{Some key problems about jet formation}

The basic questions for understanding the physics and role of jets in AGN and
$\micro$-quasars and their complex taxonomy are those related to the nature of
the constituing plasma,  their initial acceleration in the near environment of
the central black hole, their morphology as they propagate away from the 
central region, the connection of the source (disk, disk corona  or black hole
magnetosphere) and extraction of angular momentum from it. Fig. \ref{eps4}
illustrates how the plasma is extracted from the magnetized rotating  source,
spiraling and carrying roped magnetic field lines. 
  
Let us briefly recall here some points which we will not address in detail 
in the remaining part of this review.

First, it is likely that jets 
associated with quasars are a mixture of electron/proton and 
electron/positron pairs, since pure electron/positron 
jets overpredict soft X-ray radiation from quasars, while pure proton-
electron jets predict too weak nonthermal X-ray radiation 
\cite{SikoraMadejski00}. In fact, it has been suggested 
first by Sol et al. \cite{Soletal89} that jets consist of 
electron/positron pairs close to their axis, surrounded by an electron/proton 
plasma (see Fig. \ref{eps4}). 
This model has known some success since then 
\cite{PelletierSol92,BlandfordLevinson95} and  
it may naturally account for the presence of
ultra-relativistic flows in the parsec scale made of pair plasma or a mixture 
of the two while
the lobes at larger scales would be fed with the proton-electron part.
In this picture the lobes become illuminated
by the faster inner jet at some point due, for instance,
 to an instability that 
disrupts the jet. It accounts also for the tremendous 
energetic power observed in the lobes which does not need to be the
transported directly by the inner pair plasma.

Second, rotating outflows extract also rather efficiently angular 
momentum from their source allowing thus mass to accrete to the central 
object. This removal of the angular 
momentum of the accreted material is very efficient in the presence of 
the magnetic field, as is happening in stellar magnetic 
braking \cite{Shatzman}. In fact,
the magnetic lever arm $\varpi_{a}$ at the point where corotation ceases is 
larger by a factor of order 10 than the cylindrical 
radius of the footpoint of a fieldline, $\varpi_o$ (see Fig. \ref{eps7}b,
\cite{Spruit96,Livio99}). 
It is interesting that even if a tiny percentage of order of 
1$\%$ of the accreted mass rate $\dot {\cal M}_{acrr}$ is lost through a jet,
 $\dot {\cal M}_{jet}$, the major part of the angular momentum of the 
infalling gas is removed and so this gas can be freely accreted by the 
central object (see \cite{Spruit96} for a simple explanation).     
However, though angular momentum extraction by the wind can be sufficient to
account for the accretion (e.g. \cite{Ferreira97}), turbulent viscosity 
triggered by some instability can also allow for accretion. In this case, 
the outflow is likely to exist nonetheless (e.g. \cite{CasseFerreira00a}).
Thus the presence or not of a jet is certainly not an evidence of what 
really triggers the accretion, though it certainly puts some strong constraint
on it.

Let us just clearly state here that in the following sections we will mainly 
concentrate on the mechanisms which can produce {\it acceleration}, 
explain {\it collimation} and the related {\it nature} of the source 
in each case. We refer to various reviews
\cite{Spruit96,Livio99,Blandford99,Blandford00} 
where complementary issues on the problem are treated in more detail.
We are not addressing here the question of the propagation of the jet  far 
from the source, its connection to the lobes and the external medium as this 
is the subject of other reviews in this volume (e.g. Aloy) and other 
contributions of the conference.


\section{Basics of Jet Formation Theory}

\subsection{The outflow MHD equations}

Since the outflows we are describing are made of tenuous electron/proton or 
electron/positron plasma they can usually be  modeled to 
zeroth order via the set of the ideal MHD equations. Of course the collision 
rate may be so low that thermalization is not complete and each species should be 
treated separately. However, even in the well studied case of the solar wind where 
densities are probably even lower than in relativistic jets 
the fluid approach has proven to be very efficient and better 
than a pure collisionless one in describing the dynamics of the outflow. 
We shall not give here the 
general axisymmetric equations which can be found in the literature 
(as, for instance, in this volume).
Though it is an observed fact that at least part of  the
extragalactic outflows are relativistic in velocity or temperature,  it has been shown that the basic physical mechanisms at work for the
formation of  jets are the same with those operating in the classical regime, e.g.,  
\cite{Chiuehetal91,HN89}. Thus, in the following we will not 
distinguish between relativistic and non relativistic approaches, unless some 
noticeable difference exists.
 
The full relativistic set of ideal MHD equations in the 3+1 formalism can be 
found in \cite{Camenzind96} for instance and their reduction to the classical 
Newtonian limit in \cite{BreitmoserCamenzind00}. They constitute a set of highly 
nonlinear and coupled partial differential equations of the four 
spatio-temporal variables. Note that in the following we shall use 
indifferently spherical ($t$, $r$,$\theta$,$\varphi$) or cylindrical 
($t$, $\varpi$, $\varphi$, $z$) coordinates. To describe the flow one needs then to
 determine its:
\begin{itemize} 
\item mass density $\rho$, 
\item velocity field $\vec{V}$, 
\item magnetic field  $\vec{B}$ (and electric field  $\vec{E}$ 
in a relativistic treatment)
\item gas pressure $P$ (or equivalently, the  temperature $T$) of the 
fluid.
\end{itemize}
 This can be done 
by combining Maxwell's equations for the electromagnetic fields with the 
conservation of mass, momentum (Euler's equation) and energy for the 
hydrodynamic fields. The energy equation is usually (but not always) replaced 
by the simplifying assumption of a polytropic equation of state. 

Under the assumption of steadiness ($\partial/\partial t = 0$) and axisymmetry 
($\partial/\partial \varphi= 0$), the toroidal components ($B_\varphi$, 
$V_\varphi$) can be expressed in terms of the poloidal quantities
\cite{BreitmoserCamenzind00}. Simultaneously, the magnetic field on the 
poloidal plane $[r,\theta]\equiv[\varpi,z]$ (Fig. \ref{eps4}) is defined by 
means of a scalar magnetic flux function $A$,  
$\vec B_p = (\vec \nabla A \times \hat \varphi )/\varpi$ and 
the velocity field on the poloidal plane is also defined by means of 
the  mass flux function $\Psi$,  
$\vec V_p = (\vec \nabla \Psi \times \hat \varphi )/\varpi$. Note that $\Psi=
\Psi(A)$ because of the flux freezing law of ideal MHD. Practically, 
magnetic field lines and flow lines are roped on the same mass/magnetic flux
tubes as shown in Fig. \ref{eps4}. Then the momentum equation splits in the 
poloidal plane into a component along each poloidal streamline and a component across
it. Momentum balance along the poloidal flow (the Bernoulli equation) may be combined with 
momentum balance across the flow (the transfield or Grad-Shafranov equation) 
to form a system of two coupled  partial differential equations for the 
density $\rho$ and the magnetic flux function $A$. 

Irrespectively of using a polytropic equation of state between pressure and 
density, or not, this system contains integrals that depend only on the 
magnetic flux distribution, such as:
\begin{itemize}
\item the \underline{mass to magnetic flux ratio}, $\Psi_A(A) = d\Psi/dA$,  
\item the \underline{total angular momentum}, $L(A)$, 
\item the angular velocity or rotational frequency of the footpoints of 
the magnetic fieldlines anchored in the wind source, star or disk, 
$\Omega(A)$, which is also the \underline{corotation frequency}.
\end{itemize}
Note that $L/\Omega=\varpi_{a}$ must be the cylindrical radius of the field 
line $A$ at the Alfv\'enic transition in order to ensure a smooth Alfv\'enic 
transition.

If a polytropic equation of state is used, an extra conserved quantity exists 
by integrating the momentum equation along the flow: 
this is the energy per unit mass, $E(A)$, which includes kinetic energy, enthalpy,
gravity and Poynting flux. Usually the polytropic index is less 
than the adiabatic one (and even 
less than $3/2$) in order to allow for thermal acceleration \cite{Parker63}. 
This is just a way to circumvent the solution of the rather difficult problem 
of solving the full MHD equations by selfconsistently treating the heating supply in the 
plasma.  In fact even if no polytropic assumption is made, a generalized form 
of the energy conservation can be written including a heating and cooling 
along the flow \cite{ST94}. Some authors (e.g. \cite{Sakurai85}) prefer to use
the energy in the corotating frame of rotation $E'$ to show in this non 
Galilean frame explicitely the centrifugal potential. The two notations are
equivalent and $E=E'+L\Omega$ where $L\Omega$ (called the ``the energy of the magnetic 
rotator'')  is the energy a magnetic fieldline needs
to corotate at  frequency $\Omega$ and plays a crucial role in magnetic acceleration. 

The remaining part of this section, despite that equations
are not given, is rather more technical and the reader interested in the physical 
mechanisms at work may well skip it.

\subsection{Axisymmetric and time-dependent numerical simulations}

The time-dependent MHD problem has been treated only by means of 
numerical simulations for obvious technical reasons. Thus, there have been 
performed simulations of relativistic or non relativistic disk winds 
\cite{OuyedPudritz97,Krasnopolskyetal99,Ustyugovaetal99,Ustyugovaetal00}, 
outflows from a spherical magnetosphere \cite{BT99,TB00,KeppensGoedbloed00,Usmanov00}
or, from both types of sources \cite{Koideetal00a,Koideetal00b,Nobuta99,Kudohetal98}.
However, in some simulations it is not clear that the boundaries do not introduce 
spurious effects (e.g. \cite{OuyedPudritz97,Krasnopolskyetal99})
or, that the system relaxes into a reproducible final state  
(e.g. \cite{Koideetal00a,Koideetal00b,Nobuta99,Kudohetal98}). 
Another difficulty with the numerical simulations has to do with the fact 
that AGN jets often extend over lengths more than six orders of magnitude 
their width, while the available grid sizes are much smaller. 
One way out of this constraint is to solve the problem 
via a combination of numerical techniques for the near zone and analytically
solving the hyperbolic steady state problem at large distances from the 
central source (e.g. \cite{BT99,TB00}, although the very first 
accelerating region close to the base is not treated). 

\subsection{Axisymmetric and steady analytical solutions}

Several solutions of the steady MHD equations for various sets of boundary 
conditions are available analytically while there exists only one 
numerical solution for a specific and quite unique set of boundary conditions obtained  
by Sakurai \cite{Sakurai85} for stellar winds showing very weak collimation
(i.e. logarithmically) around the rotational axis.
Basically the main difficulty is the fact that the set 
of the steady  and axisymmetric MHD equations are 
of mixed elliptic/hyperbolic type, as opposed to the hyperbolic by nature
time-dependent equations. Then, from the 
causality principle, a physically acceptable solution needs to cross three
critical surfaces: the slow magnetosonic, the Alfv\'enic and the fast
magnetosonic surfaces. However, the exact positioning of those critical 
surfaces is not known {\it a priori}  but is only determined simultaneously
 with the solution. It is for this reason that only a few classes of such 
exact MHD solutions have been studied so far. They
can be obtained by employing a separation of the variables ($r, \theta$) in the
poloidal plane (see Vlahakis and Tsinganos \cite{VT98} for a general technique to 
obtain such solutions)  combined with a suitable choice of the MHD integrals 
$\Psi_A$, $L(A)$, $\Omega (A)$ and $E(A)$. 
It is worth to note that this systematic construction unifies all existing 
analytical models of cosmic outflows, such as the classical Parker wind 
\cite{Parker63}
and the Blandford and Payne disk wind \cite{BP82}, in addition to 
uncovering new and interesting global models \cite{VT98}. 
The best studied classes of such solutions are characterized by   
radial and meridional self-similar symmetries because all 
quantities scale with the spherical radius $r$ or the colatitude $\theta$
respectively. 

The {\it first} family with radially self-similar symmetry is appropriate 
to winds emerging from disks (e.g. 
\cite{BB78,BP82,ContopoulosLovelace94,Li96,Leryetal99,CasseFerreira00b,Vlahakisetal00,VT98}
 and references therein). A relativistic extension of these self-similar models
 exists (see e.g. \cite{Lietal92}) although by dropping one essential element: gravity. 
No intrinsic scale length exists in this case and 
all quantities  scale as a power law of the radius, similarly to the Keplerian
law for the velocity in the disk. The key assumptions in this class of 
solutions are that the poloidal Alfv\'en Mach number $M$ and the cylindrical 
radius $\varpi$ of a particular poloidal fieldline $A$={\it const.}, in units 
of the cylindrical radius $\varpi_{a}$ at the Alfv\'en point along 
the same poloidal fieldline, are solely functions of the colatitude $\theta$.
In this case surfaces of constant $M$ are assumed to be conical, and the critical
surfaces too.

The {\it second} family is characterized by the meridional self-similar 
symmetry and is appropriate to winds emerging from a spherical source, 
although the physical variables are not spherically symmetric and the boundary
conditions are functions also of the 
colatitude (\cite{TTS97,STT99,VT98} and references therein). 
Even though these solutions seem to be more natural to describe 
stellar winds, they do not exclude the presence of a surrounding accretion 
disk and they can on an equal footing describe a quasi spherical 
corona or magnetosphere around the central object. 
The key assumptions also in this case are that the poloidal Alfv\'en Mach 
number  $M$ and the cylindrical radius $\varpi$ of a particular poloidal 
fieldline $A$={\it const.} are solely functions of the spherical radius $r$.
In this case surfaces of constant $M$ are assumed to be spherical and so are 
the critical surfaces.

\subsection{Boundary conditions and singularities}

As we mentioned in the previous section, for the construction 
of a steady solution one has to carefully cross the 
appropriate critical surfaces  encountered at the characteristic MHD speeds,  
corresponding to the three MHD waves 
propagating in the medium (but not to the elliptic/hyperbolic transitions 
as we discuss in the following). It effectively results in reducing the 
number of free boundary conditions \cite{Bogovalov97}. 
Note that in the relativistic case the number of critical surfaces is the 
same with the nonrelativistic case because the light cylinder singularity is 
combined with the Alfv\'enic one \cite{BreitmoserCamenzind00}. As a corollary, 
this generalized  Alfv\'enic singularity reduces to the classical one in the 
non relativistic regime and to the light cylinder when the mass loading is 
negligible.

\begin{figure}[b]
\begin{center}
\includegraphics[width=.85\textwidth]{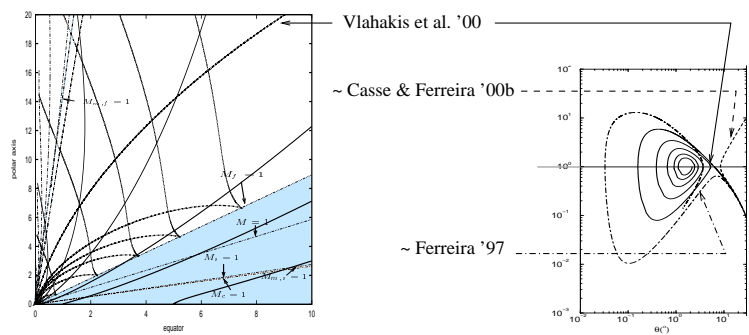}
\end{center}
\caption[]{Topology of an analytical disk wind solution in the region of the
fast magnetosonic transition in the [$M_f$,$\theta$] plane. Examples of three 
solutions of this type are drawn:
one that crosses the critical point and two which do not.}
\label{eps6}
\end{figure}
It is conventional to associate the crossing of 
the slow surface to fixing the mass loss rate and the crossing of the Alfv\'en
to fixing the magnetic torque \cite{Spruit96}. 
The crossing of the last  singularity usually remains
more subtle but is essential to ensure that no instability or shock with 
the external medium is going to propagate backward and possibly changes the
structure of the solution. This is well known for the Parker solar wind,
where terminal shocks will naturally make breeze solutions to evolve into 
the wind solution. 

To illustrate this point we show in Fig. \ref{eps6}
three disk wind solutions belonging to the radial self-similar class and the 
topology of
another typical set of such solutions around the fast  surface in the plane of
colatitude and fast magnetosonic Mach number (instead of the Mach number for a 
classical hydrodynamic wind). A careful look at the
behaviour of those solutions shows that one is crossing exactly the fast 
magnetosonic transition \cite{Vlahakisetal00} while the other two behave close to
the critical surface either like a breeze solution \cite{Ferreira97} or
like a terminated solution \cite{CasseFerreira00b} in Parker's terminology for
the solar wind, as illustrated on Fig. \ref{eps6}.  Although they differ 
in the way they connect to the underlying disk, they basically show very 
similar properties. This indicates that, by tuning the heating deposition along
the flow or the polytropic index, a physical connection with the disk and the
crossing of the fast point is possible despite it has not been done yet. It
also suggests that the crossing of the last surface is not so crucial for the
connection with the disk but it validates, nonetheless, the widespread use of such 
disk wind solutions.
In fact all solutions terminate after some point due to the presence of a 
spiral singularity as shown in Fig. \ref{eps6}, which makes even more crucial the
presence of a shock not only from observational arguments but also from 
theoretical ones.

\begin{figure}[b]
\begin{center}
\includegraphics[width=.85\textwidth]{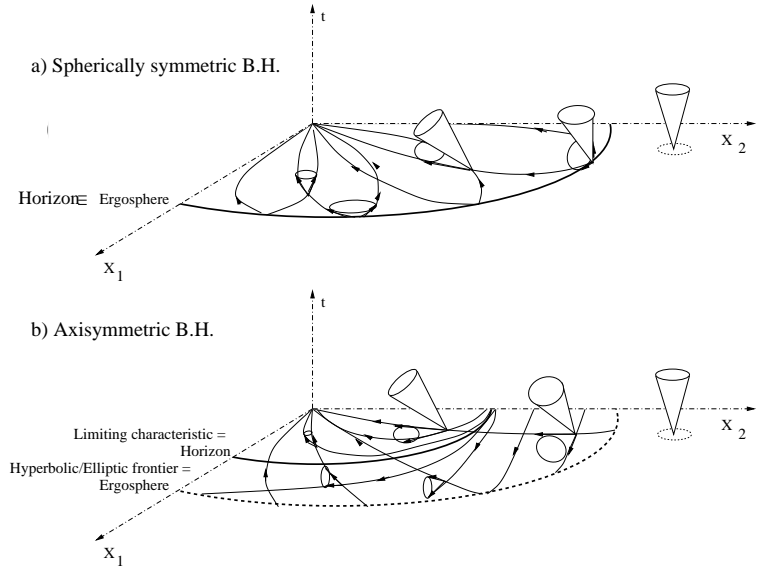}
\end{center}
\caption[]{Ergosphere, horizon and light cones around a black hole after
Carter \cite{Carter72}. Time is along the vertical axis and space in
 the horizontal plane $[X1,X2]$. In a) sketch of a spherically symmetric black hole
and in b) a rotating axisymmetric one.}
\label{eps5}
\end{figure}

\subsection{Singularities and horizons}

We discuss now the true nature of the 
singularities, pointing out that a strict analogy between MHD signal propagation 
and  the light propagation  in the neighborhood of black holes (see  Carter \cite{Carter72})
exists.

{\it First}, as it is well known in the theory of steady and spherically symmetric 
black holes (Schwarschild black holes in the case of vacuum), the horizon where
no information or light signal can escape coincides with the ergosphere where
the system of the relativistic equations changes nature, from elliptic outside the 
horizon to hyperbolic inside it. 
As illustrated in Fig. \ref{eps5}a the system is hyperbolic in time and information 
emitted at the speed of light propagates along a ``cone'', the light-cone. 
In a steady state and far from the black hole, 
the light-cone's projection is a circle and thus information can propagate
in all directions of the $X_1$ - $X_2$ plane, similarly to water waves on a static pool.
  $X_1$ and $X_2$ are two coordinates in space. 
The equations are elliptic. But because of gravity, light 
is deflected and inside the horizon, equations change to hyperbolic, the cone 
projection gives characteristics, as the trail of a boat, and information can 
propagate only inside these. Furthermore, all characteristics converge to the 
center of the black hole such that no information can escape (Fig. \ref{eps5}a). 

The same is true and well known in MHD outflows where light is replaced by MHD 
waves and the spatial coordinates are reversed ($X_1$ corresponding to 
$1/X_1$) such that the center of the black hole becomes the asymptotic outer
connection of the MHD outflow with the extragalactic medium.
In spherically symmetric outflows (or equivalently when the poloidal
geometry of the flow is fixed),  each of the three singularities coincides in fact
with a spherical surface which marks the transition in the nature of the 
equations from hyperbolic to elliptic\footnote{There is an extra transition 
at the cusp velocity which is not a singularity while the Alfv\'en singularity is 
in a parabolic domain but we shall not enter here in these details.}.

{\it Second}, if the black hole is rotating, the event horizon and the 
ergosphere where the system changes from elliptic to hyperbolic split. 
It is easy to understand this physically (see Fig. \ref{eps5}b). As the
geodesics rotate, the light cone first inclines itself such that beyond
the ergosphere characteristics appears. One of the family of characteristics
still connects with the external medium so it is still possible to propagate 
information backward. Once inside the real horizon the light cone is inclined 
and directed towards the center. At this point the outer space is causally 
disconnected from the interior of the horizon. 

The analogy between the event horizon and the MHD singularities in terms of
limiting characteristics or separatrices has been recently recognized \cite{Bogovalov97}
and illustrated explicitly by examples of self-similar solutions \cite{Tsinganosetal96}.
Note that for the slow magnetosonic horizon the situation is a bit more 
complicated because 
the slow waves have triangular wave fronts instead of circular \cite{Vlahakis98}. 
However, the ultimate horizon is the one associated with the fast waves. 
In addition, the so-called ``classical'' critical points correspond to the 
ergosphere where the equations change from elliptic to hyperbolic. This 
point has been underestimated. In fact, it is known for the ergosphere of black
holes that it can appear as a singularity but a suitable choice of the Killing 
vectors eliminates it. The same must be true for the ``classical'' critical 
points where the poloidal velocity equals one of the wave speeds.

{\it Third}, the analogy can still be pushed one step further. There are 
only two cases where the horizons can be defined locally, either, for 
static black holes where the horizon coincides with the ergosphere (Fig. \ref{eps5}a), 
which corresponds to the spherical Schwarzschild solution in vacuum, or, in the
circularity limit where the horizon coincides with the rotosurface,
which means that there is only rotation and no convection.  In all the
other cases, the horizon can be defined only globally that is as a limit
of the characteristics, once the solution of the metrics is known (i.e. as a 
limiting characteristic). 

The circularity limit for the vacuum solution corresponds to the
Kerr rotating black hole (Fig. \ref{eps5}b). In this special case, one can construct the
solution \cite{Carter72}, by means of a separation of the variables $r$
and $\theta$ while $t$ and $\varphi$ are ignorable. The $\theta$ component can
be solved by using Legendre's polynomials. This way of constructing the global 
solution is identical to the one used for self-similar flows. In both
cases the form of the singular surfaces is known a priori and this is
definitely NOT the self-similar assumption that ``modify'' the critical
points as we have written for so long.

\underbar{The first conclusion} is that in the more general axisymmetric 
case, there is little hope that we can determine the limiting characteristics 
{\it a priori} in MHD outflows. However, these are the true singularities of the flow.

\underbar{The second conclusion} is that this problem of horizons concerns not 
only steady solutions but also numerical time-dependent solutions. Even if they
do not appear as real mathematical singularities, they must be present in the
sense that characteristics on the outer boundaries should all be directed
outwards, which has not been the case of all simulations as we already 
mentionned.


\section{Acceleration}

Once valid solutions of the outflow equations are obtained, either numerically
or analytically, we may study the physical mechanisms that accelerate and collimate 
the outflow transforming it from a wind to a jet. From a rather general perspective, an 
observational characteristic of many cosmic plasma outflows 
is that they seem to be {\it accelerated} to relatively high speeds which may 
even  reach values close to the speed of light in the most powerful AGN jets. 
The most often invoked mechanisms to accelerate these flows  are of 
thermal, or, of magnetocentrifugal origin. We shall discuss first in the following 
magnetocentrifugal acceleration since it is widely considered as the most relevant 
mechanism for the acceleration of AGN jets. 

\begin{figure}[b]
\begin{center}
\includegraphics[width=.45\textwidth]{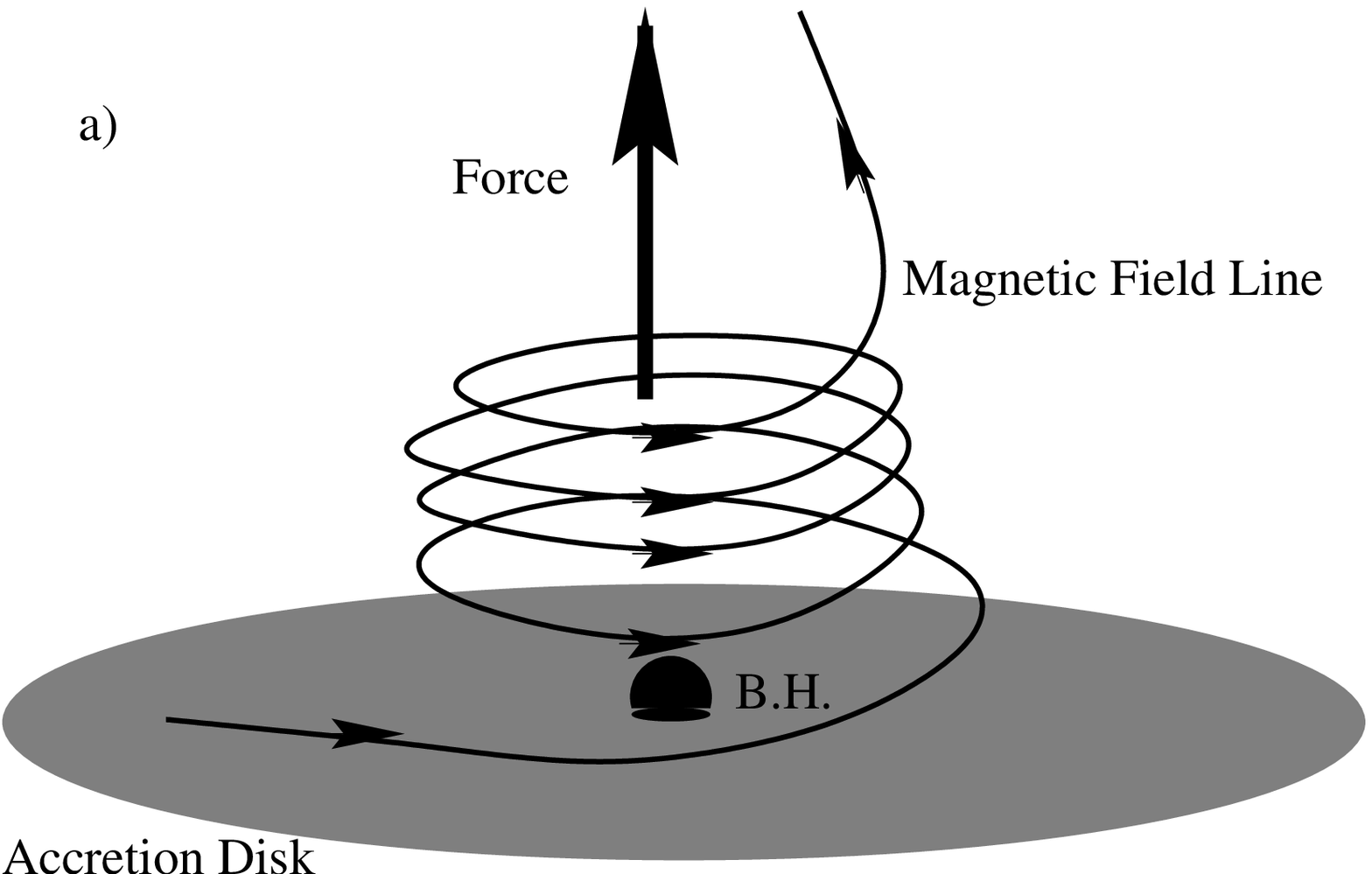}
\includegraphics[width=.45\textwidth]{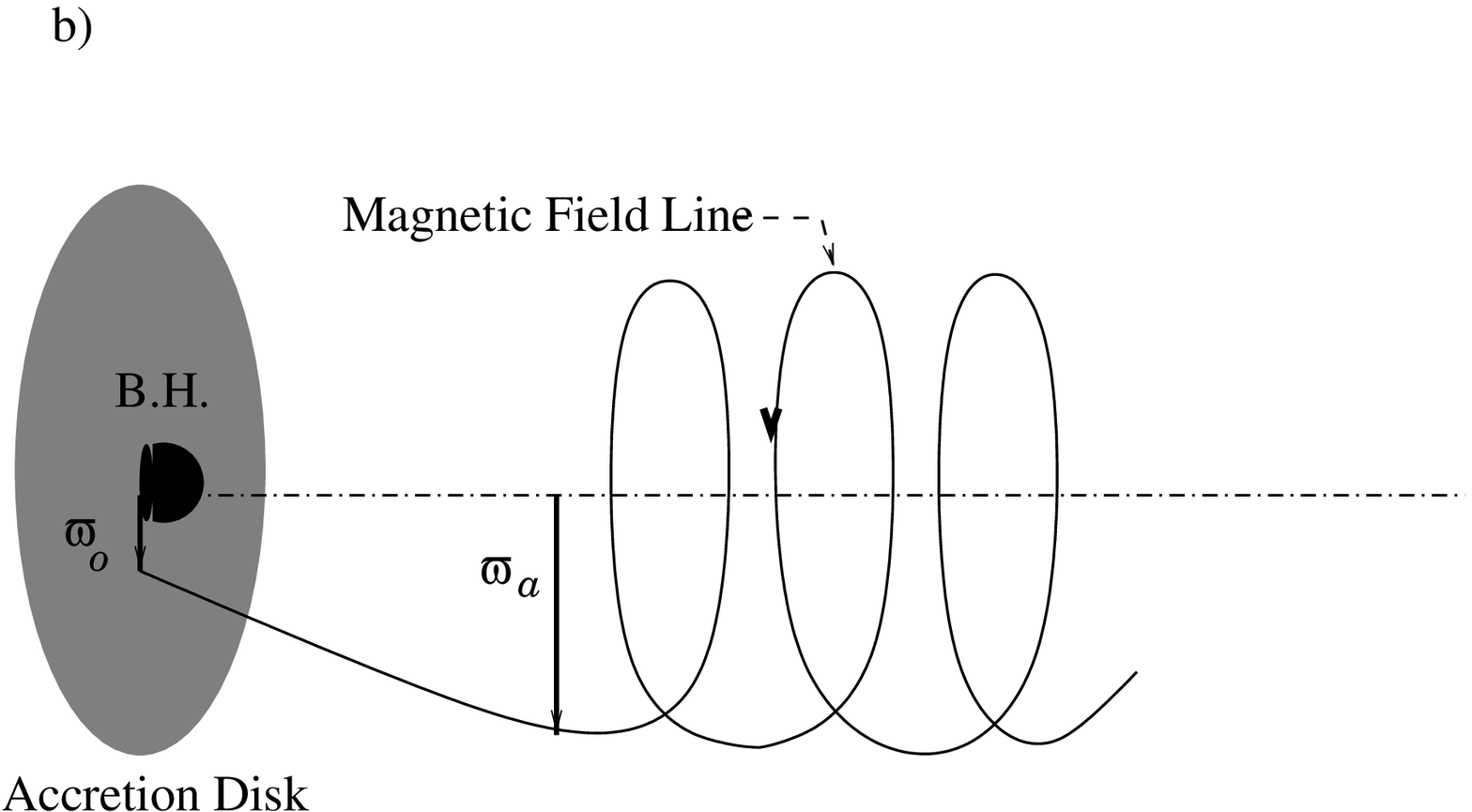}
\end{center}
\caption[]{{\bf a)} Acceleration by toroidal (azimuthal) magnetic pressure. 
The fieldline is wounded by rotation and acts as an uncoiling spring. 
{\bf b)} Magnetocentrifugally driven wind. The acceleration is similar to that 
of a `bead on a wire' and it operates from the disk footpoint $\varpi_o$ up to the 
magnetic lever arm $\varpi_a$ where corotation stops. 
Further downstream the magnetic field is rapidly wound up and magnetic collimation 
is obtained because of the pinching magnetic tension (after Spruit \cite{Spruit96}).}
\label{eps7}
\end{figure}
\subsection{Toroidal magnetic field acceleration}

The simplest magnetic driving mechanism is the so-called 
'uncoiling spring' model (Uchida and Shibata 1985, \cite{UchidaShibata85,Kudohetal98}) or 
`plasma gun' \cite{Contopoulos95} where a toroidal magnetic pressure
 builds up due to the rotation of the fieldlines which are anchored in the disk. 
Evidently, there is a net force pushing the plasma upwards, as shown in Fig. \ref{eps7}a. 
This mechanism is mainly seen in numerical simulations (e.g., \cite{Shibata90}). 
After the initial transient phase when a torsional 
Alfv\'en wave develops and drives the initial acceleration of the flow, the 
solutions  converge to a weakly collimated structure, where the confinement is done 
by the toroidal magnetic field. However, for the numerical constraints we mentioned 
above, the outflow cannot be simulated in regions far from the base to follow 
realistically its degree of collimation. Second, such numerical simulations were
able to follow the jet for one or two rotations around the central body. Hence, such a 
mechanism seems to be at work only to explain intermittent ejection, something 
equivalent to Coronal Mass Ejections (CMEs) in the solar wind that 
travel on top of a global more steady structure.

It is interesting that instabilities by a spiral wave in the disk producing an 
Alfv\'en wave  have been also advocated to explain the intermittent ejection from 
$\micro$Quasars\cite{TaggerPellat99}. These instabilities can be at work to explain
various features, as we briefly shall discuss later and transients on a time scale 
of a few rotations around the black hole. However, we need to model at the same time 
some more continuous ejection, similarly to the solar wind where despite instabilities
and CMEs, there always exists a steady wind outflow. 

\subsection{Magneto-centrifugal acceleration from the disk}

In order to accelerate jets magnetically, the most popular scenario 
is the magnetocentrifugal acceleration  from an accretion disk via the classical 
Blandford \& Payne \cite{BP82} acceleration mechanism. In this case, the plasma 
consists probably mostly of electrons/protons, either relativistic or not. 
Electrons/positrons could be accelerated in the same way but it seems more difficult to
produce them above the keplerian disk than in or close to the black hole's magnetosphere for 
energetic reasons, e.g. \cite{HenriPelletier91}. As the 
poloidal magnetic field is dominant up to the Alfv\'en radius, it practically enforces 
an approximate plasma corotation similarly to a 'bead on a rotating wire' 
(see Fig. \ref{eps7}b). In the corotating frame there is then a centrifugal potential
which accelerates the flow outwards provided that the line is sufficiently inclined 
($\theta > 30^o$ from the pole for a cold non relativistic plasma). 
Note that this condition is  less restrictive if there is some heating or the 
gas is relativistic \cite{SpruitCao94,Spruit96}. 
This point should not confuse the reader, because the centrifugal 
force does not alone accelerate the flow but the opposite and it is its combination
with a strong poloidal $\vec{B}$ field that allows for the acceleration. Eventually the
acceleration comes from the conversion of Poynting energy flux to kinetic energy 
flux. The gain in kinetic energy is proportional to the energy that brings the 
magnetic field lines into rotation, i.e., the energy of the magnetic rotator
$L \Omega$. Such a magnetocentrifugal driving mechanism seems to be efficient in 
disk winds wherein a hot corona is not an absolute requirement. 

Besides a long list of self-similar models (see 
\cite{VT98,KoniglKartje94,KoniglWardle96,Ostriker97}), 
the same mechanism has shown
to be successful in various numerical simulations \cite{OuyedPudritz97,Krasnopolskyetal99}.

This mechanism has some limitations. For example, it requires high
magnetic field strengths at the disk level (which have not been measured so 
far) and 
also a large magnetic lever arm is needed in order to obtain a terminal speed which is a 
few times the Keplerian speed ($\lesssim 10^5$ km/s). 
Moreover, only a very small fraction of the accreted mass can be 
ejected at very high speeds, once a connection with a realistic disk structure is properly made 
\cite{Ferreira97,Li96,CasseFerreira00a}. By realistic disk structure we mean that in the accretion 
disk resistivity, turbulence and viscosity are taken into account within the 
hypothesis that the disk is pervaded by a large scale mean magnetic field. However, the 
presence of a hot corona on top of the disk is enough to eliminate this limitation 
\cite{CasseFerreira00b}. If these conclusions hold unchanged in the 
presence of a local dynamo and/or disordered magnetic field is not clear yet
\cite{HeinzBegelman99,Blandford00}. 

\subsection{Magneto-centrifugal acceleration from the black hole magnetosphere}

A wind outflow could also be extracted ``magneto-centrifugally'' from a black
hole's magnetosphere through the  Blandford \& Znajek mechanism 
\cite{BlandfordZnajek77} (see also \cite{Blandford00} for  more references). 
First, the rotational energy of the accreting 
black hole is extracted by a large scale magnetic field accreted onto the black 
hole, then converted to Poynting flux and finally to relativistic 
electron/positron pairs. As far as the plasma itself is concerned there is no 
difference with the previous mechanism, since it is ultimately the Poynting flux that 
accelerates it. However, there are two basic differences. First, the Poynting flux 
is extracted from the black hole. This is physically consistent since some 
angular momentum of the opposite sign is lost simultaneously into the black hole. 
Second, there is enough energy in the magnetosphere to produce electrons/positron 
pairs, such that this mechanism favors leptonic ejection. 
It has been argued that pairs would suffer Compton drag; however, this may be 
a real problem for radiatively driven winds but not if there is an extra mechanism 
to accelerate the flow sufficiently to overcome these radiative losses.

The efficiency of this mechanism has been recently put into question by several
authors (see for instance \cite{Livioetal99}). The main argument is that the 
extraction of energy from the black hole through this process is at most as 
efficient as the extraction of energy from the disk. This only means that the two mechanisms
are likely to operate simultaneously. Again this could very well be in favour of
a leptonic jet or beam extracted from the back hole embedded in a hadronic heavy 
wind/jet coming from the disk \cite{Soletal89,HenriPelletier91}. Moreover this does
not apply necessarily to the gamma ray emission which may still get its energy from 
the black hole \cite{Blandford00}.

\subsection{Radiative acceleration}

The first alternative to a magnetically driven wind is a radiatively driven wind. 
For instance, in the 'Compton rocket' model a disk produces electron/positron 
pairs which are accelerated by the radiation produced by the 
annihilation of this plasma. Although Compton radiative losses exist they
are not sufficient to prevent completely the plasma acceleration \cite{ODell81,ChengODell81}.

Radiatively driven models for disk winds  of electrons/protons have been proposed 
with a radiative pressure due to dust or, due to lines coming either from the disk or 
the central source (see \cite{Progaetal00} and further references in the introduction). 
The key point in such models is that they result in at most a few tens of 
thousands of km/s for the outflow speed ($\sim 50,000$ km/s). 
This may be enough to explain most of the winds from radio quiet AGN, but it is unable to 
explain the acceleration of the powerful jets associated with radio loud AGN, or, the 
mildly relativistic flows seen in some radio quiet AGN. 
It is thus likely that in those objects, radiative acceleration may operate as a minor 
contribution or, in combination with other mechanisms, like magnetocentrifugal driving 
\cite{KoniglKartje94}.

 Alternatively, it has been shown that a magnetized 
cloud  of relativistic electrons can be radiatively driven up to relativistic
velocities if the interaction between photons and  particles comes through a synchrotron process 
\cite{ghis90,ghis92}. However the cross section of this mechanism is critically dependent
on the geometry of the interaction and some assumptions  are necessary `a priori' for the treatment
of the equations. 
 
\subsection{``Thermal'' acceleration}

The other alternative to magnetic acceleration is classical thermal driving, as it is the 
case in the low- and high-speed solar wind where the heating and part of the
pressure is provided by the dissipation of acoustic waves, electric currents, etc., 
or more efficiently by Alfv\'en waves. 
 In this case the presence of a hot corona around the disk and/or the magnetosphere
is essential for the acceleration, which  
is proportional to the sound speed, i.e., to the square root of the coronal 
temperature. 

Then, in a rather crude estimate, if the $10^6$ K corona produces a thermally 
driven wind with a terminal speed around 300 km/s, a corona with a temperature of $10^9$ K 
for both the ions and the electrons could result in a terminal speed around 10,000 km/s.  
On the other hand, if $10^9$ K is the temperature of the electrons while the temperature 
of the protons is $10^{12}$ K, a wind results with a terminal speed of the order of the speed 
of light, 300,000  km/s. 
Of course at this point relativistic effects should be taken into account properly. For 
ultra-relativistic flows the adiabatic sound speed is only  $c/\sqrt{3}$; 
however, this is without taking into account the existence of extended heating in the corona.
Most of all the heating by waves, in particular (torsional) Alfv\'en waves, could be 
very efficient. For instance, it
has  prooven to be efficient enough, even with small amplitudes, to explain the 800  km/s
of the fast solar wind \cite{Usmanov00}. By extrapolation, we may guess that it should be
able to produce very high speeds in AGN outflows. Waves of large amplitudes could be even
more efficient, producing turbulence, and this is not very different from the transient
torsional Alfv\'en ``wave'' seen in numerical simulations \cite{Kudohetal98}, except
that the production of such waves should be continuous. 

In fact, this mechanism combined with magnetocentrifugal driving, has found some
success in the literature \cite{Livio99}, like for  
the acceleration in the corona from ADIOS \cite{BlandfordBegelman99}, from a Keplerian 
disk \cite{CasseFerreira00b}, from a black hole magnetosphere  through a shock
\cite{Das00,Koideetal00a,Koideetal00b} or, more in general, from any kind of spherical 
corona \cite{ST94,TTS97,STT99}.

Thus, thermal acceleration in a broad sense is likely to be as efficient as the magnetic
processes. We could then suggest that both may be at work in disk winds for
electron/proton plasmas (where it also allows to have higher mass loss rates,  
as we already mentioned \cite{CasseFerreira00b}) while the electron/positron pairs 
would be more likely magnetically driven from a black hole magnetosphere.
This seems also to be suggested by recent numerical simulations 
\cite{Koideetal00a,Koideetal00b} in which two flow streams  are accelerated near 
the black hole after passing through some shock in the accretion disk. 
Despite the fact that the disk is governed by ideal
MHD and it is not clear if the simulations can hold more than one or two
rotation times, it is worth to note that these simulations show a double component of 
the wind, a inner magnetocentrifugally driven part and an external pressure driven 
component. 


\section{Collimation}

Once the outflows are accelerated, they will propagate in the form of either 
collimated beams or uncollimated winds. However, apart for the case of the solar wind, 
uncollimated flows are hardly observable, while jets are observed in 
several astrophysical environments, from star formation regions to distant 
AGN. This is mainly due to the much higher density inside jets as opposed to that in  
loosely collimated winds. Furthermore, in radio-loud AGN where  beams move at relativistic 
speeds, the emission may be largely amplified by Doppler boosting if the jet is pointed 
towards the observer, which is probably not the case is radio-quiet sources.
Again the two basic mechanisms responsible for collimation may be of thermal or magnetic
origin.
 
\subsection{Pressure confinement}

An outflow is thermally confined if the surrounding medium has a higher 
pressure than the flow, such that there is a pressure gradient forcing the outflow to 
collimate along its ejection axis. In other words, only outflows underpressured with 
respect to their surrounding environment may be thermally confined. 
In fact, such a situation seems to  occur in many extragalactic jets, as deduced 
from X-ray data implying a hot plasma surrounding early-type 
galaxies and clusters of galaxies \cite{Ferettietal95}.  

The 'twin exhaust model'' based on an analogy with the De Laval Nozzle, was the 
first effort to thermally confine jets \cite{BlandfordRees74}. 
However, this confining mechanism has been by now excluded because 
it requires the throat of the nozzle to be located rather far from the central 
object, as it works for both collimating and accelerating the flow. 
To remedy that, the idea of an external medium  but only collimating the outflow 
 has  been suggested \cite{FabianRees95}.
Meridionally self similar models \cite{TTS97,STT99} have shown that 
cylindrical collimation could arise naturally from inward pressure forces
but with some contribution by the magnetic field too. 
 In an pure thermally collimated flow at the end the jet should collapse onto the
rotational axis, unless it rotates fast enough such that the centrifugal force may counteract 
the external pressure. However, the strength of rotation is likely to be related to the strength of
the  magnetic field (because of dynamo for instance) such that they are usually low simultaneously.

So thermal confinement may play a role in FR I and Seyfert types of AGN but
probably not in FR II which are known to have a very poor environment so 
there cannot definitely be a unique mechanism for all AGN.

\subsection{Poloidal magnetic confinement and subfast flows}

With toroidal magnetic fields known to be unstable in tokamaks, 
it has been suggested that toroidal confinement and pinching should be unstable,
and collimation could be achieved by poloidal magnetic fields alone \cite{SpruitCao94}.  
This is supported by the parallel magnetic field measured at small parsec 
scales in some extragalactic jets \cite{Spruit96}. However recent observations of highly 
optically-polarized compact radio-loud quasars (HPQ) has shown that the electric 
vectors of the polarized 43 GHz radio cores are roughly aligned with the inner jet 
direction indicating magnetic fields perpendicular to the flow  \cite{ListerSmith00}. 
Magnetic fields are also known on larger scales to be perpendicular to the jet axis in 
FR II sources while they are parallel to the jet axis in many FR I sources.  
However parallel does not necessarily mean that it is not helicoidal and there is no 
toroidal field. The parallelism  could simply be due to a 
strong velocity shear across the jet's cross section as it is explained in \cite{Blandford00}.

Poloidal magnetic fields can induce some mild collimation in the outflow 
in the region from the disk up to the Alfv\'en transition (see Fig. \ref{eps7}b) where the 
plasma in the transfield equation governing the morphology of the flow is basically 
dominated by magnetic forces.
Beyond this distance, the jet becomes superAlfv\'enic and the hydrodynamics of the flow overcomes 
magnetic forces, in such a way that collimation will stop. To continue  poloidal collimation
on large distances, the jet must remain subalfv\'enic. But then, the jet will be very sensitive
to shocks and instabilities that can propagate upstream from far distances and destroy
the whole equilibrium. 

Similar problems occur in asymptotically cylindrical solutions 
\cite{ContopoulosLovelace94,Ostriker97} of radially self-similar disk-wind models. 
Subfast outflows as those proposed by Ostriker \cite{Ostriker97} attain only low Alfv\'en Mach 
numbers and such solutions are structurally unstable \cite{Vlahakisetal00}. In fact, as we already 
wrote, all the other solutions of those models are terminated because of the spiral singularity
(see Sec. 2.4).

\subsection{Toroidal magnetic confinement and stability}

Another confining mechanism, which is in fact supported by observations 
\cite{ListerSmith00}, is the magnetic confinement 
of the outflow by a toroidal magnetic field wound around the jet, the
so-called hoop-stress paradigm \cite{Spruit96}. This mechanism works both for 
under- and over-pressured jets. Observations of perpendicular magnetic fields 
\cite{ListerSmith00} imply that such beams carry some electric current that eventually 
closes  at their surface or outside. Note that the building of the toroidal magnetic 
field is done at the expense of the Poynting flux. Thus, in a pure magnetic jet all the 
Poynting flux cannot be converted to kinetic energy, if part of it remains to confine 
the jet. Obviously reality in most cases, and particularly in the case of AGN, may 
involve a combination of thermal and magnetic processes in the acceleration and 
confinement of the outflow.

Despite the observations that we mention above, there is still a vigorous debate on 
whether magnetic
instabilities may ultimately disrupt the jet, or not. In particular, it has been shown in the
context of a pure magnetic jet without rotation that instabilities are always present
\cite{Begelman98}. Rotation in a pure hydrodynamic flow is also known  
to have a destabilizing nature and hydrodynamic
instabilities may also disrupt the jet \cite{Bodoetal98}, though relativistic
jets are more stable (see Aloy this volume). However, the combination of
toroidal magnetic fields and rotation is more subtle, since the two ingredients 
act in opposite directions. 

Recent numerical simulations and extended analytical work \cite{KimOstriker00} gave 
support to the result that magneto-rotational instabilities may develop rapidly. 
Such instabilities for a cold plasma tend to favor the formation of an inner denser core 
in a jet. 
Similar results have been obtained from a local analysis of the ballooning modes 
\cite{Kersaleetal00}, though there is no precise calculation of any growth rate. 
In particular, the inner part of the jet with a vanishing current density on the polar 
axis, is particularly unstable to magnetic shearing in this analysis.

On the other hand, a non linear analysis of current driven instabilities
 (\cite{Leryetal00b} and ref. therein) has shown that the instability instead of 
disrupting the jet leads to a reorganisation of the current density. Instabilities in
more complex jets from Keplerian disks have also been studied \cite{Leryetal00a} and
present interesting structures, again forming a dense core jet surrounded by a return 
current in a cocoon.
 
Parallely, magnetorotational instabilities have also been studied using the flux tube 
approximation 
\cite{Hanaszetal00}, which is likely to give the most unstable mode, including only
toroidal fields but with all the other ingredients, such as shearing or buoyancy. This 
study agrees with the fact that jets with a Keplerian velocity profile, as well as 
the outer region where the jet connects to the external medium are subject to strong 
instabilities. Conversely, the inner parts of a jet can be completely stabilized for
a flat, or a solid rotation profile, provided -- and this does not appear in the studies
mentioned above -- that there is an increase of the density away from the axis.
In this case jets from the central source, like those obtained in meridional self similar
models, should be more stable than jets from Keplerian disks. It also explains the edge 
brightening seen in some sources, because the instabilities are likely to occur mostly at 
the edges.
Eventually instabilities are sources of reacceleration in the plasma and ultimately 
radiation. Finally, it suggests that hollow jets (not empty jets!) should be more 
stable than dense core ones.

Altogether, then it follows from both,  observations and theoretical studies, that 
one better be careful before claiming that toroidal instabilities will disrupt 
rotating MHD jets. However,  instabilities do in general
exist and they are obviously a crucial element for explaining a few structures
we see in both observed jets and simulations. The issue is rather crucial 
and is a subject by itself in this volume (see the article by Aloy). 

\subsection{Asymptotic equilibria}

With hoop stress and pressure gradients collimating winds into jets, it
is interesting to wonder which kind of asymptotics the outflow takes. 
A useful general analysis for magnetically dominated flows has been performed by 
Heyvaerts \& Norman \cite{HN89}, generalized to relativistic flows in  
\cite{Chiuehetal91}. 

In the case where pressure and centrifugal forces drop asymptotically faster than 
the magnetic forces do, then the final equilibrium state should be force-free, 
and the asymptotical morphology of the flow is related to the electric current 
flowing. To summarize, a given flux tube collimates to
\begin{itemize} 
\item {\it cylindrical} asymptotics if there is a net poloidal current spread in it;
\item {\it paraboloidal} asymptotics if the net poloidal current is zero;
\item {\it radial} asymptotics if there is a net current but it flows inside the region 
of the radial asymptots. 
\end{itemize}
Of course, the whole jet could be in principle cylindrical if the return current lies 
outside the jet or, in a current sheet (cf. observations \cite{ListerSmith00}). Note that
flows with return current sheets are known to exist and the best example where it is
observed and measured {\it in situ} is the solar wind. However from the analysis of 
Heyvaerts \& Norman (1989) the possibility of mixed asymptotics with cylinders surrounded 
by cones ($=$radial asymptotics) is not excluded.
Asymptotic solutions of such flows have been successfully
constructed especially in the context of relativistic flows either for pure cylindrical 
asymptots \cite{Fendtetal95}) or mixed radial and cylindrical (e.g. \cite{Nitta97},\cite{BT99}).

An analogous analysis to \cite{HN89} has been recently proposed,  which nevertheless 
arrives to opposite conclusions \cite{Okamoto99}. It extends the study to the 
quasi-asymptotic domain (called asymptotics), i.e. for
$z\lesssim\infty$ and connects the curvature of the streamlines to the direction  of the 
current density. It is shown that the collimating part of the outflow is related to 
the enclosed current, while the outside region of return current should cause the outflow 
to decollimate. However, it is postulated that cylindrical and radial asymptotics are not 
accepted as a valid possibility in MHD outflows because of their ``violation of causality'' 
and because cylinders correspond to a specific direction, as claimed therein. 
Instead, a continuous deflection towards the polar axis or the equatorial
plane is preferred. However, this inevitably leads to an inconsistently infinite density 
there while close to the polar axis the underlying assumptions are not valid any longer as we 
explain below. A more interesting relativistic generalisation of these results is obtained in  
\cite{BeskinOkamoto00} where collimation is not rejected as a possibility 
but it is pointed out that the presence of decollimated flows could explain strong 
equatorial flows seen around several compact objects. 

All previous results rely on the strong hypotheses that pressure and centrifugal 
forces drop asymptotically. In order to have this, two assumptions are
made. First, that the cylindrical radius of any flux tube is assumed to be much larger than 
its value at the Alfv\'enic transition $\varpi/\varpi_a \rightarrow \infty$ -- this is 
not necessarily true for cylindrical asymptots even for  $\varpi \gg \varpi_a$. Second, 
it is assumed that we do not remain too close to the polar axis \cite{Okamoto99}. 
More general asymptotics have been found \cite{ST94,TTS97,STT99} including pressure and 
rotation
but using the assumption of meridional self-similarity which cannot hold
indefinitely far from the axis (where the return current is well known to exist for 
instance). Cylindrical and radial asymptots are found in agreement with the Heyvaerts \&
Norman \cite{HN89} conclusions despite the different assumptions.
 
Then, after considering the asymptotic behaviour in general, one needs to connect it to 
the source \cite{Fendtetal95,Leryetal98} and if possible by solving 
selfconsistently the transfield equation \cite{ST94,TTS97,VT98,STT99,BT99,TB00}, 
a topic we take up in the next section.


\section{On a possible classification of AGN}

\subsection{An energetic criterion for the collimation of outflows}

Apparently a missing parameter in the vertical classification of AGN in 
Fig.~\ref{eps2} corresponds to a variation in the degree of collimation, going from
the winds of Seyferts, to the jets from FR I and then to the powerful jets from FR II 
type of AGN.
Thus we need a criterion for collimation to get a quantitative information on how much
the flow will expand. 

In fact several models have found a `fastness parameter' $\alpha$ given by
\begin{equation}
\label{fastness}
\alpha^2 = \frac{L\Omega}{V_a^2}
\,,
\end{equation}
where $L\Omega$ is again the energy of the magnetic rotator and $V_a$ the poloidal Alfv\'en
velocity at the Alfv\'en transition where $\varpi=\varpi_a$. This parameter was originally
introduced by Michel \cite{Michel69} to measure the fastness of the magnetic rotator which 
accelerates the flow of a cold plasma. In an equatorial wind, when this energy dominates 
we have a fast magnetic rotator and the wind is magneto-centrifugally driven. Conversely, 
when thermal acceleration is dominant the magnetic rotator is termed slow. 
It appeared that $\alpha$  also controls the degree of collimation 
in several analytical models (e.g. Ferreira \cite{Ferreira97}, L\'ery et al.
\cite{Leryetal98}).
In the numerical approach followed by Bogovalov \& Tsinganos \cite{BT99,TB00}
the degree of collimation is determined by a similar parameter $\alpha$   
expressing the ratio of the angular velocity times the Alfv\'en spherical distance to 
the initial constant speed of an initially nonrotating split-monopole type of a magnetosphere. 

In fact in most of these models boundary conditions were exactly spherically symmetric
on the source except for rotation \cite{Leryetal98,BT99} or, the magnetocentrifugal forces
 were dominant in collimating and accelerating  the flow \cite{Ferreira97}. If gas pressure
gets important and/or the boundary conditions in density are not spherically symmetric,
there seems to be some changes in the degree of collimation,  although the role of the fastness
parameter remains qualitatively the same \cite{TB00,CasseFerreira00b}. 
In the meridionally 
self-similar approach followed by Tsinganos et al. \cite{ST94,TTS97,VT98,STT99} the degree of
collimation is not only related to the fastness parameter but also to the distribution of
the thermal content. In particular,  specific criteria for the collimation of winds were
derived in the frame of these models \cite{ST94,STT99} that we shall summarize here.
\begin{figure}[b]
\begin{center}
\includegraphics[width=.9\textwidth]{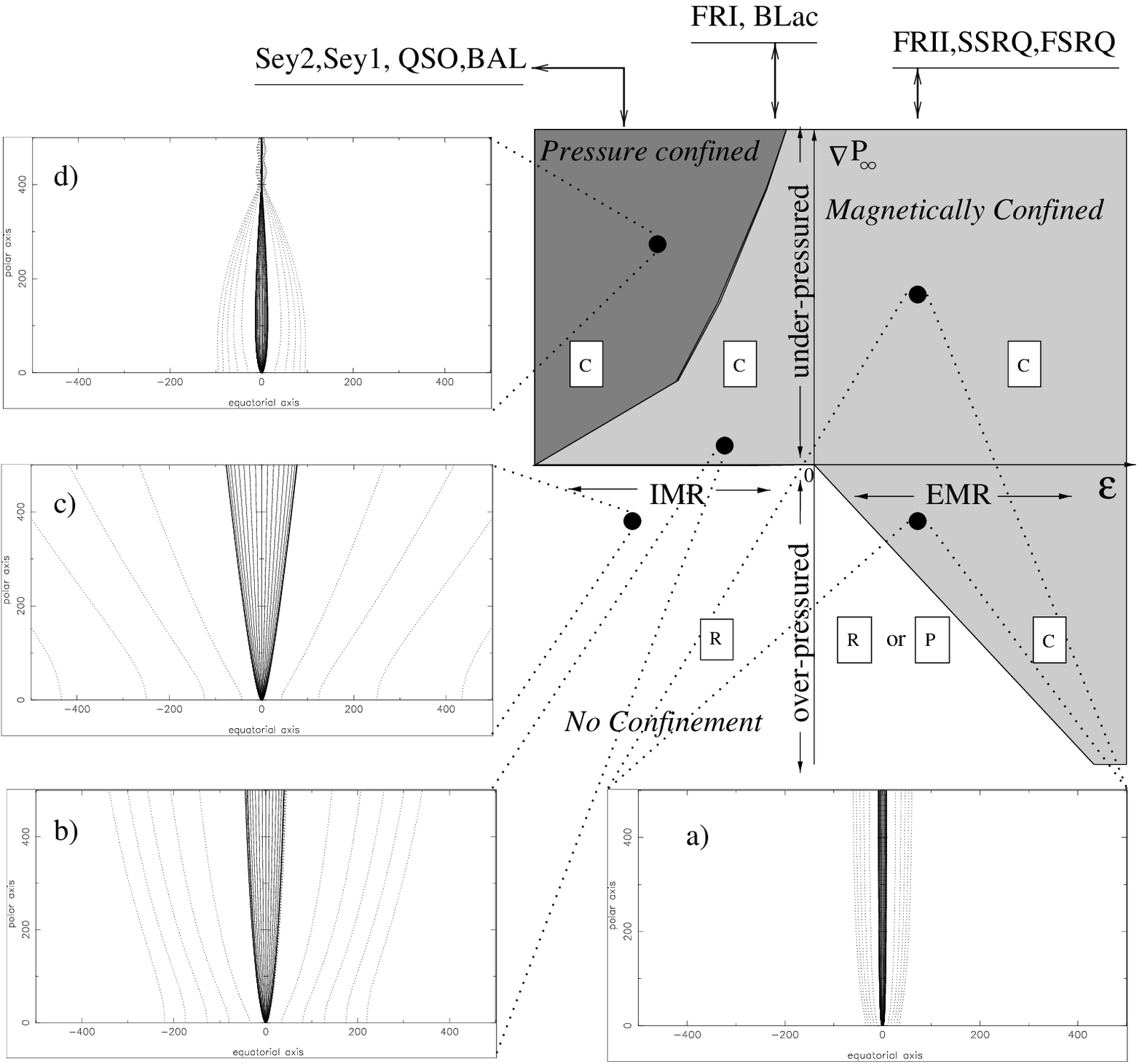}
\end{center}
\caption[]{Degree of collimation obtained  as a function of
the asymptotic transverse pressure gradient (vertical
axis) and the efficiency of the magnetic rotator (horizontal axis)
with typical solutions for winds in (b) and  (c), jets in (a) and 
stopped jets in (d). EMR are on the left, IMR on
the right, underpressured jets on top and overpressured flows
below. ``R'' corresponds to the domain or radial asymptotics, ``P'' to
paraboloidal ones and ``C'' to cylindrical.
See text for details. After Sauty, Tsinganos and Trussoni 
\cite{STT99}.}
\label{eps8}
\end{figure}

 Usually in an outflow the thermal input in the form of internal energy and 
external heating is not all fully converted into other energy forms: unless the 
terminal temperature is zero, there remains some asymptotic thermal content, 
$h (\infty , A )$. By subtracting from the total energy  $E(A )$ the heat content at infinity,
 $h (\infty , A )$, we obtain a new streamline constant, 
$\tilde E (A )$, which will be the total \underbar{convertable} specific energy 
along the given streamline $A$, i.e., the energy which can be converted to other forms. 
Finally, the volumetric total convertable energy is $\rho (r, A ) \tilde E( A )$. 

It turns out that in meridionally self-similar flows the difference of the 
volumetric convertable energy between a nonpolar streamline and a polar streamline 
normalized to the volumetric energy of the magnetic 
rotator $\rho (r, A ) L(A ) \Omega (A )$ is a constant $\epsilon^{\prime}$ 
\cite{STT99}, 
\begin{equation}
\label{epsilonprime}
\epsilon^{\prime } = 
\frac{
\rho (r, A ) \tilde E(A ) - \rho (r, {\rm pole} ) \tilde E( {\rm pole} )} 
{\rho (r, A ) L (A ) \Omega (A )}
\,. 
\end{equation}

This quantity $\epsilon^{\prime}$ plays a crucial role in the asymptotic 
shape of the streamlines, in the sense that the necessary condition for 
cylindrical asymptotics is $\epsilon^{\prime}>0$. In other words, 
cylindrical collimation is controlled by a 
single parameter, $\epsilon^{\prime}$, representing the variation 
between a fieldline $A$ and the pole, of the sum of the (volumetric) 
poloidal and toroidal kinetic energies, the Poynting flux, the gravitational 
potential and the converted thermal content. 

If everything is homogeneous in the medium except rotation and the energy of
the magnetic rotator, which always increases away from the axis, then this
parameter is very similar to the fastness parameter ($\epsilon^{\prime }\sim
\alpha^2$). Conversely, if the density increases also substantially away from the 
axis while temperature drops so much that the thermal heating cannot lift the plasma, 
acceleration will be done at the expense of the Poynting flux. Thus the Poynting
flux won't be any longer available to collimate the flow and the degree of 
collimation will decrease.

In the specific model we are describing, this can be put in a more quantitative
form because this parameter splits into two terms \cite{STT99}:
\begin{equation}
\label{ee}
\epsilon^{\prime} \equiv \mu + \epsilon
\,.
\end{equation}
$\mu$ represents the variation across the streamlines of the 
thermal content that is finally converted into kinetic energy and gives a measure 
of the thermal pressure efficiency to collimate the outflow. 
$\epsilon$ is the efficiency of the magnetic rotator to collimate.

In fact $\mu$ can be written
\begin{equation}
\mu = \frac{P(r,A)-P(r,{\rm pole})}{P(r, {\rm pole})}
{V_\infty^2\over V_*^2} 
\,,
\end{equation}
where $V_{\infty}$ and $V_*$ are the polar asymptotic and Alfv\'en speeds, 
and $P(r, A)$ the pressure along the streamline $A$. 
For under-pressured flows ($\mu>0$) the pressure gradient force is outwards helping 
collimation. Conversely, over-pressured jets ($\mu<0$) and 
iso-pressured jets ($\mu=0$) can collimate only magnetically.

On the other hand, the parameter $\epsilon$ is equal to the excess of the magnetorotational 
energy on a nonpolar  streamline which is not used to drive the flow, in units of the energy 
of the magnetic rotator. It can be evaluated at the base of the flow $r_o$,
\begin{equation}
\epsilon ={L\Omega - E_{{\rm R},o}+\Delta E_{\rm G}^* \over E_{\rm {MR}} }
\,.
\end{equation}
with
\begin{equation} 
 \Delta E_{\rm G}^* 
= - {{\cal G}{\cal M} \over r_o}
\left[  1-{T_o(\alpha)\over T_o({\rm{pole}})} \right]
\,,
\end{equation}
where ${\cal G}$ is the gravitational constant, ${\cal M}$ the central mass
and $T_o$ the temperature at the base of the flow $r_o$.
The energy of the magnetic rotator $\Omega L$ is mainly stored in the form 
of Poynting flux, i.e. $E_{{\rm R},o}$ (the rotational energy) is usually a negligible quantity 
in the above  expression. In other words, $\epsilon$ measures
how much of the  energy of the magnetic rotator is not used to escape the
gravitational well and is available for magnetic collimation alone.   
If there is an excess of this energy on non polar 
streamlines, magnetic forces can collimate the wind into a jet.
Thus, when $\epsilon>0$ we have an {\it Efficient Magnetic Rotator 
(EMR)} to magnetically collimate the outflow into a jet, and an {\it Inefficient 
Magnetic Rotator (IMR)} if $\epsilon\le 0$ \cite{STT99}.

Results are summarized in Fig. \ref{eps8} in the parameter space with typical solutions
represented in the poloidal plane. For each solution,  the lines are simply a cut in this 
plane of the magnetic flux tubes, i.e. a projection in this plane of the wounded 
streamlines as in Fig. \ref{eps4}. First we see that jets
from EMR are very well collimated independently of being under- or over- pressured, 
as illustrated with the solution in Fig. \ref{eps8}a. For IMR the situation is more
complex. If the flow is iso- or over-pressured it cannot collimate so it is
conical with an asymptotic vanishing pressure as shown in Fig. \ref{eps8}c. 
Even if it is under-pressured at the base but with vanishing pressure 
or becoming over-pressured asymptotically as shown in Fig. \ref{eps8}b
the cylindrical collimation is very loose because it is 
due only to the presence of a weak magnetic 
field. Conversely, if the pressure remains strong all the way it can refocalize strongly
the jet and squize it as in Fig. \ref{eps8}d). In fact this last situation looks like
a stopped jet somehow.

\subsection{Application to the classification}

We found that the asymptotic morphology of the outflow is controlled by
the efficiency of the magnetic rotator and the pressure gradient across the streamlines, 
Eq. (\ref{ee}). The same model shows interesting jet solutions in relation with
the various flows seen in AGN (Fig. \ref{eps8}), so we can try to use it in understanding their
taxonomy. 
 The efficiency of the magnetic rotator is related to the magnetic properties of the 
central object in the AGN and/or its disk while the pressure gradient is
related to the pressure variation across the streamlines. 
We may assume that this can be somehow related to the environment 
through which the jet propagates (this is of course an extra assumption to the model
itself). Thus, we may discuss the following possibilities in the framework of the
classification scheme of AGN shown in Fig. \ref{eps2}, as they are summarized 
in Table 1 (this scheme is also shown on the top of Fig \ref{eps8}).
In this classification, we move from Type 2 to Type 0 because of orientation
effects, as we already discussed. The new interesting element is that a  
classification from one class to another results now as the efficiency of the magnetic
rotator and the environment change.

\begin{table}
\caption{AGN classification according to orientation 
and efficiency of magnetic rotator} 
\begin{center}
\renewcommand{\arraystretch}{1.4}
\setlength\tabcolsep{5pt}
\begin{tabular}{lllll}
\hline\noalign{\smallskip}
Radio-emission (Type: 2, 1, 0)  & Magnetic Rotator, 
$\epsilon$ & Collimation, $\epsilon^\prime$ \\
\noalign{\smallskip}
\hline
\noalign{\smallskip}
Quiet (Sey.2, Sey.1, BAL \& QSO) & Inefficient, 
$\epsilon \ll 0$ & Weak, $\epsilon^\prime \lesssim 0$ \\
Loud  (FR I, BL Lac) &  Intermediate effic., & Good, $\epsilon^\prime \gtrsim 0$  \\
& $\epsilon \sim 0$ &   \\
Loud  (FR II, BLRG \& SSRQ, FSRQ) &  Efficient, $\epsilon > 0$ 
& Tight, $\epsilon^\prime >0$   \\
\hline
\end{tabular}
\end{center}
\label{Tab1a}
\end{table}
                                     
(I). {\it \underbar{Inefficient} magnetic rotators, $\epsilon \lesssim 0$, 
corresponding to radio-quiet AGN (Seyferts, etc\dots) with uncollimated or, loosely 
collimated outflows.}    
One possibility is that $\epsilon^{\prime} < 0$ such that the AGN produces 
a radially expanding outflow (Fig. \ref{eps8}c). This may happen if the source is in a rich 
environment such that latitudinally  we have an over-pressured outflow $\mu>0$. 
The other possibility is that $\epsilon^{\prime}\gtrsim 0$, i.e. 
$\epsilon^{\prime}$ is marginally positive such that the AGN produces a `weakly' 
collimated  jet, i.e., collimation occurs slowly at large distances (Fig. \ref{eps8}b).  
This may happen, for instance, if we have a latitudinally under-pressured 
outflow, $\mu >0$. 
Hence,  if the central source is an IMR ($\epsilon \lesssim 0$) the density drops 
quite rapidly with the radial distance and this could be related to the weaker 
outflows in  Seyfert 1 and 2 galaxies and radio-quiet QSO's. We cannot exclude 
that in this case if pressure does not drop rapidly enough we have a stopped jet 
like in (Fig. \ref{eps8}d) with a shock at the refocalizing point, as it has been
suggested for Seyferts (e.g. \cite{Rosso94}).

(II). {\it \underbar{Efficient} magnetic rotators, $\epsilon > 0$, 
corresponding to radio-loud AGN of high luminosity with well collimated and powerful
jets (FRII, etc\dots).}   
In this case, since $\epsilon$ obtains high positive values, we have a tightly 
collimated jet (Fig. \ref{eps8}a), regardless of the value of $\mu$, 
i.e., regardless if the jet propagates in a rich or poor environment.

(III). {\it \underbar{Intermediate efficiency} magnetic rotators, 
$\epsilon^{\prime} > 0$, 
corresponding to radio-loud AGN of low luminosity (FRI, etc\dots) with collimated jets.} 
The two possibilities are either that $\epsilon$ is marginally positive and 
$\mu <0$, or, $\epsilon$ is marginally negative and $\mu >0$.   
In this case, we always have $\epsilon'>0$ and 
hence the outflows always have asymptotically cylindrical flux tubes.
Note also that many extragalactic jets, as deduced from X-ray data on the 
hot surrounding plasma, seem to be propagating in rich environments 
\cite{Ferettietal95}. For example, this seems to be the case with FRI type of 
Radio Galaxies \cite{PrestagePeacock88}.   

If the strength of the magnetic rotator reduces, one expects a smooth 
transition from a jet to a loosely collimated wind and finally to a radial wind. 
This would correspond to moving from the radio-loud quasars and Blazars 
of Fig. \ref{eps2} to the radio-quiet Seyfert galaxies and QSO's. The same transition
would be true if the closeby environment, possibly the corona of the central engine, 
becomes more and more dense. This is also consistent 
with the different kinds of parent galaxies: early-type (with very low density interstellar gas) 
for radio-loud AGN, Blazars, QSO,  and spiral galaxies for Seyferts. 

Despite that the model 
on which our conclusions are based is clearly 
nonrelativistic, we conjecture that its basic trends should be preserved in 
relativistic cases as well if we rely on previous relativistic extensions of non 
relativistic results \cite{Chiuehetal91}. However, collimation of relativistic winds from a 
spherical source (see \cite{BT99}) seems  rather difficult to achieve 
because the plasma has a very high effective density in this case
(something that somehow fits into the criterion for collimation we previously 
discussed). This can be solved in two ways. Either, the jet is launched almost along its rotational
axis as it is usually done in numerical simulations of relativistic disk winds, or, 
there is in fact an indication that the relativistic pair plasma beam is confined by an
 heavier more extended hadronic component which is not relativistic or only mildly (cf. 
\cite{HenriPelletier91}).

\section{Concluding remarks}

In this review we started with a long catalogue of the taxonomy of outflows from 
AGN and $\micro$Quasars but soon we realized that all such outflows share common 
characteristics that can be understood in physical terms. 

First, magnetic collimation of winds into jets appears to be a rather general 
property of the MHD equations governing plasma outflows (the hoop stress
paradigm) and is likely to survive to the instabilities of the toroidal magnetic field,
although such instabilities indeed must be present and can deeply modify the 
morphology of the outflow. 
Pressure gradients also contribute to confine the outflows in addition to toroidal 
magnetic fields.  
In extragalactic jets and on scales of several kpc, pressure confinement by the environment 
seems to be present especially in the less luminous  and less collimated ones, 
while close to the center the jet may be either magnetically or thermally confined.

Second, the transformation of magnetocentrifugal energy into kinetic energy seems to
be a natural driving mechanism for outflows from black hole magnetospheres 
and accretion disks, but the presence of very hot coronae in AGN 
and the necessity along the rotational axis to have a thermal driving
indicates that the contribution from the thermal 
energy is essential if not dominant, with appropriate heating processes
occurring in the plasma. 

And third, the jet composition is likely to be made of both electron/positron pairs and 
electrons/protons, the first being more likely to be extracted from the central 
magnetosphere while the second from the more extended corona and surrounding disk.

The MHD acceleration/collimation mechanisms can work in very different 
astrophysical scenarios, whenever we have a rotating magnetized body 
such as a supermassive black hole surrounded by 
an accretion disk.  Even though the basic thermal/magnetic 
driving and confining mechanisms discussed here should be qualitatively 
valid also for relativistic velocities, more detailed modeling of such 
relativistic jets from AGN is needed at this point. However a 
consistent modeling of jets in AGN requires the crossing of all MHD singularities.
And we have seen how difficult it is to solve the steady equations and how
important it is for putting carefully the boundary conditions in numerical
simulations. At this point the analogy of horizon/limiting characteristics 
and ergosphere/elliptic-hyperbolic-transitions may be of some help in the
future as there is already a long experience in numerical simulations of
black holes and how to tackle with this difficulty.

Finally, we have reviewed the standard unification scheme of AGN. The fact that
some classes of objects transform into other classes with the viewing angle, 
seems to be basically secure by now. And, in this article we added that there is a 
physical criterium separating the various classes among themselves. 
We see that the degree of collimation does not depend only on the spin of the
central black hole or the fueling or the composition of the environment, but
in a subtle composition of all these processes which can be expressed in terms
of the energetic distribution.
In this sense it allows to reconcile the different scenarios  
proposed to explain the taxonomy of winds and jets around compact objects,
in particular the FRII/FRI dichotomy.


%

\end{document}